\def\={\hbox to 0pt{\hss$=$\hss}}
\def\beq{\begin{equation}}
\def\eeq{\end{equation}}
\def\beeq{\begin{eqnarray}}
\def\beeqn{\begin{eqnarray*}}
\def\eeeq{\end{eqnarray}}
\def\eeeqn{\end{eqnarray*}}
\def\re#1{(\ref{#1})}
\def\wG{\widetilde G}
\def\hG{\widehat{G}}
\def\p{\partial}
\def\pdx{{\partial \over \partial x}}
\def\pdof#1{{\partial #1\over \partial x}}
\def\pd#1#2{{\partial^{#1} #2\over \partial x^{#1}}}
\def\sijk{\sum_{\{i,j,k\}\atop \epsilon \,C_3}}
\def\Lm{\Lambda}

\documentclass[12pt]{article}

\begin{document}

\begin{titlepage}
\vspace*{-2cm}
\begin{flushright}MUM-THP 98/001 \\ January 1998 \\
    (Revised April 2000)
\end{flushright}
\vskip1.4cm
\begin{center}
{\Large\bf Exact renormalization group flow equations  
\\[.2cm] for free energies and $N$-point functions
\\[.3cm] in uniform external fields}
\end{center}
\vskip1.5cm
\begin{center}
{\bf Geoffrey R. Golner}\footnote{E-mail: ggolner@mum.edu}
\end{center}
\begin{center}
\vskip.3cm 
{\it Department of Physics
 \\ Maharishi University of Management
 \\ Fairfield, IA 52557, USA}
\vskip1.5cm
\bf Abstract
\end{center} 
 \begin{quote}
    
    We project the Wilson/Polchinski renormalization group
     equation onto its uniform external field dependent effective free
     energy and connected Green's functions.  The result is a
     hierarchy of equations which admits a choice of ``natural"
     truncation and closure schemes for nonperturbative approximate
     solution.  In this way approximation schemes can be generated
     which avoid power series expansions in either fields or momenta. 
     When following one closure scheme the lowest order equation is
     the mean field approximation, while another closure scheme gives
     the ``local potential approximation."  Extension of these closure
     schemes to higher orders leads to interesting new questions
     regarding truncation schemes and the convergence of
     nonperturbative approximations.  One scheme, based on a novel
     ``momentum cluster decomposition" of the connected Green's
     functions, seems to offer new possibilities for accurate
     nonperturbative successive appproximation.
    \end{quote}
\vfill
\end{titlepage}

\baselineskip=0.65cm

\section{Introduction} 

In recent years there has been a resurgence of
interest in applying Wilson's exact renormalization group (ERG)
equation \cite{wilsonkogut} to problems in quantum field theory. 
Starting with Polchinski \cite{polchinski}, who derived Wilson's ERG
equation in a context familiar to field theorists, it was first used
to provide improved, simplified proofs of perturbative
renormalizability for many field theoretic systems \cite{pertren}. 
More recently, others have used this equation, or its Legendre
transformed version \cite{wett,morijmp}, to study, in various
nonperturbative approximations, the qualitative, and in some cases
quantitative, behaviors of many different systems, including scalar
\cite{scalar1,scalar2} and vector \cite{nvecLPA,nvect} field theories,
gauge theories \cite{gauge}, fermionic systems \cite{fermi}, and many
phenomenological issues such as top quark and higgs masses
\cite{elltop}, bound states \cite{ellbound}, and various aspects of
QCD \cite{qcd}.  Recent reviews may be found in
\cite{scalar1,morrev,nato,bagnuls}.  

The challenge of nonperturbative
approximations has been to reduce the infinite dimensional space of
couplings generated by the ERG equation to some computationally
manageable size.  This is often initially done simply by truncating
the space to a few operators (usually the relevant and marginal
operators of the canonical theory), and useful qualitative and even
quantitative results are obtained by this approach.  Improved
approximations can then be sought by enlarging the space of effective
interactions in some systematic way with the hope that results will
converge towards the exact result.  However, without a small
parameter, as in perturbation theory, to give some measure of control
over the degree of approximation, the degree of ``improvement" is
dificult to evaluate.  In addition, additional nonphysical solutions
are often generated as the finite space of couplings is enlarged
\cite{fermi,nonphys,morconv}, making it often difficult to identify
with any {\it a priori} justification the physical solution sought. 

Another approach, which from the start includes an infinite number of
interactions, is the derivative expansion \cite{golner,morderiv,ball},
essentially a functional power series expansion of the Wilson
effective action in powers of momenta coupling the Fourier transformed
field variables.  In this way all powers of the field are included at
each level of approximation.  Its lowest order of approximation, known
as the ``local potential approximation" (LPA)
\cite{scalar2,LPArev,aoki}, is quite successful in providing
qualitative and even reasonably good quantitative information
regarding phase diagrams and critical exponents.  It appears to
generate no spurious solutions and permit the entire space of
candidate renormalized theories to be searched for solutions in a
simple, systematic way \cite{morrev,morconv,morderiv,morsol}.  Its
second order of approximation gives impressive accuracy when compared
with other approaches ($\epsilon$-expansion, exact, and high
temperature expansion) \cite{golner,morderiv}, and has offered the
hope of leading to a practical method of successive approximation.  

A third approach, which has not been much used outside of perturbation
theory, is complementary to the derivative expansion in that it
involves an expansion of the Wilson effective action in powers of the
field operators \cite{wilsonkogut}, the expansion coefficients being
the (zero-field) connected Green's functions of the theory.  Thus, all
powers of momenta are included at each stage of approximation.  This
approach has provided an alternative route to the $\epsilon$-expansion
for scalar field theories \cite{shuklagreen} as well as, in the
Legendre transformed version, a promising initial nonperturbative
exploration of bound state questions \cite{ellbound}.  

In all of the above approaches the question of convergence of the
approximation method involved is dificult to address {\it a priori}
\cite{morijmp,morderiv}, and although the approximations often improve
when taken to their {\it next} higher order (usually the only one
practical to compute, so far), there are indications that this good
fortune cannot continue indefinitely.  In \cite{morconv} Morris
studies the convergence of an expansion of the LPA in powers of the
field variable.  He demonstrates that the exact solution of the LPA
equation has singularities in the complex plane that prevent the
convergence of the power series expansion.  If the presence of such
singularities is a general feature of the ERG, one can expect it to
prevent convergence of any expansion that involves a power series in
the field operators.  This would eliminate the hopes of convergence of
the first and third methods discussed above.  The derivative expansion
also appears to have its own problem with convergence.  In
\cite{dunne} Dunne provides an exact solution for the effective
potential for QED$_{2+1}$ in a particular inhomogeneous external
magnetic field.  From this he derives an all orders derivative
expansion and shows that it is asymptotic.  We take this to suggest
that power series expansions in either fields or momenta are unable to
provide a convergent approximation scheme for the ERG equation.

It is interesting to contrast this situation with that of lattice
real-space RG approximation schemes \cite{lattice}, where a reliable
(apparently convergent) method of successive approximation seems to
exist.  One orders the space of interactions in terms of the
interaction's relative degree of locality, {\it i.e.}, nearest
neighbor, next nearest neighbor, {\it etc.}, and truncates
accordingly.  This locality property does not seem to be readily
applicable to momentum-space formulations; it certainly has no
relation to the power series approximations discussed above, as all
powers of both the field and momenta are typical components of local
lattice interactions.

In this paper, motivated by the above considerations, we attempt to
construct a method of successive approximation for Wilson/Polchinski's
ERG equation for a scalar field theory in $D$ dimensions that is not
based on power series expansions.  In particular, each order of
approximation involves all powers of the field and (beyond the lowest
order) all powers of momenta.  In the final analysis it becomes
essentially an expansion in the {\it number} of momenta coupling the
Fourier transformed field variables, though it is, strictly speaking,
in the space of connected Green's functions rather than the space of
Wilson effective interactions that this expansion takes place.  It is
accomplished by projecting the ERG equation onto its uniform external
field dependent effective free energy and connected Green's functions,
creating an infinite hierarchy of partial differential equations
(PDEs) to be truncated and solved numerically.  A set of ``fluctuation
relations" (derived below) allows a choice of truncation schemes to be
investigated.  One of these schemes, based on a novel ``momentum
cluster decomposition" of the connected Green's functions, appears to
have many advantages over the standard approach.  Although it still
remains to be numerically investigated, it appears to be a likely
candidate for a convergent method of successive approximation as well
as offering new possibilities for extending the power of the
renormalization group approach in many directions.  

The remainder of this paper proceeds as follows.  In Section 2 we
introduce our system and notation and give a summary of the essential
steps of the derivation of the Polchinski ERG equation and its
Legendre transformed version.  In Section 3 we derive a rescaled
version of the Polchinski equation that is particularly well suited
for the analysis that follows.  We also show that it exactly
reproduces Wilson's rescaled ERG equation.  In Section 4 we derive a
formalism for expressing the connected Green's functions in terms of
the solution of the rescaled ERG equation.  We demonstrate that our
rescaling procedure gives the correct scaling laws for connected
Green's functions.  We also derive the set of ``fluctuation relations"
which will prove important in our later sections.  In Section 5 we
derive flow equations for the connected Green's functions and present
an analysis that suggests that the ``obvious" truncation strategy will
not be a convergent method of successive approximation.  In Section 6
we present a new solution strategy based upon the ``fluctuation
relations" presented in Section 4.  We define new ``$N$-point momentum
clusters" and derive ``momentum cluster flow equations" based upon a
``momentum cluster decomposition" of the original connected Green's
($N$-point) functions.  We compare these new equations with the
equations of Section 5 and discuss the apparent advantages of this new
formulation.  In Section 7 we offer some concluding comments.

One final note: Although in this paper we
treat the calculation of the Gibbs free energy and connected Green's
functions via the Wilson/Polchinski ERG equation, our entire approach
can be equally applied to the calculation of the effective potential
and one-particle-irreducible (1PI) vertex functions via the Legendre
transformed ERG equation \cite{wett,morijmp}.  To reflect this more
general utility we use the more general terminology ``$N$-point
functions" in the title and section headings of this paper. 

\section{Flow Equations} 
\label{sec:flow} 

Our system is a
self-interacting scalar field, $\phi (x)$, coupled to an external
source, $J(x)$, in $D$ Euclidean dimensions.  We obtain the connected
Green's functions from the $IR$-cutoff, $UV$-regulated generating
functional $W^{\Lm_0}_{\Lm}[J]$ evaluated via the Wilson/Polchinski
renormalization group equation.  The derivation of this equation and
of its Legendre transformed counterpart is, by now, fairly standard
\cite{morijmp,ellbound}.  For completeness, and to establish our
notation, we will simply summarize, following Ellwanger
\cite{ellbound}, the essential steps.  

The bare theory is regularized
in the $UV$ via a cutoff $\Lm_0$ in the propagator.  We further
introduce a running infrared cutoff, $\Lm$, and corresponding
propagator \beq P^{\Lm_0}_\Lm(q^2)\equiv (R^{\Lm_0}_\Lm(q^2))^{-1}
=\frac{K_{\Lm_0}(q^2)-K_{\Lm}(q^2)}{q^2+m^2} \eeq with
\[K_{\Lm}(q^2)\to 1\hspace {0.3in}\rm {for}\hspace {0.3in}q^2\ll
\Lm^2\] \beq K_{\Lm}(q^2)\to 0\hspace {0.3in}\rm {for}\hspace
{0.3in}q^2\gg \Lm^2.  
\eeq 

The generating functional for cutoff connected Green's functions can
be represented as the functional integral \begin{equation}\label{gf}
\exp\bigl(-W^{\Lm_0}_{\Lm}[J]\bigr) =\int{\cal D}\phi\,\exp\biggl\{
-{1\over2}(\phi ,R^{\Lm_0}_{\Lm}\phi )
-S_{\mathrm{int}}^{\Lm_{0}}[\phi]+(J,\phi )\biggr\}, \end{equation}
where $(J,\phi)\equiv \int_q J_q\phi_{-q}$, $\int_q\equiv
{1\over(2\pi)^D} \int {\mathrm{d}}^D q$, and $\phi_q$ and $J_q$ are
the Fourier transforms of $\phi(x)$ and $J(x)$ respectively.  Here the
kinetic term of the bare action is represented in terms of the inverse
propagator $R^{\Lm_0}_{\Lm}(q^2)$, with the remaining part of the bare
action denoted by $S_{\mathrm{int}}^{\Lm_0}[\phi]$.

The result of the functional integration in \re{gf} may be formally
represented by \beq \label{rep2} {\mathrm{e}}^{-W^{\Lm_0}_{\Lm}[J]}=
{\mathrm{e}}^{{1\over2}(J,P^{\Lm_0}_{\Lm}
J)}{\mathrm{e}}^{D^{\Lm_0}_{\Lm}}
{\mathrm{e}}^{-S_{\mathrm{int}}^{\Lm_0}[\phi]}
\Bigm|_{\phi=P^{\Lm_0}_{\Lm} J} \eeq with \beq \label{dop}
D^{\Lm_0}_{\Lm}={1\over2}\biggl(P^{\Lm_0}_{\Lm}{\delta
\over\delta\phi}, {\delta \over\delta\phi}\biggr).  \eeq We define an
effective interaction $S_{\mathrm{int}}[\phi,\Lm]$ by \beq \label{ei}
{\mathrm{e}}^{-S_{\mathrm{int}}[\phi,\Lm]}=
{\mathrm{e}}^{D^{\Lm_0}_\Lm}
{\mathrm{e}}^{-S_{\mathrm{int}}^{\Lm_0}[\phi]}.  \eeq We obtain
Polchinski's flow equation for $S_{\mathrm{int}}[\phi,\Lm]$ by
differentiating equation~\re{ei} with respect to $\Lm$: \beq
\label{polch} {\p S_{\mathrm{int}}\over\p\Lm} ={1\over2}\int_q{\p
P^{\Lm_0}_\Lm (q^2)\over\p\Lm}
\biggl\{{\delta^2S_{\mathrm{int}}\over\delta\phi_q\,\delta\phi_{-q}}
-{\delta S_{\mathrm{int}}\over\delta\phi_q}{\delta
S_{\mathrm{int}}\over\delta\phi_{-q}}\biggr\}.  \eeq Our initial
condition is
$S_{\mathrm{int}}[\phi,\Lm_0]=S^{\Lm_0}_{\mathrm{int}}[\phi]$.

The generating functional for cutoff connected Green's functions,
$W^{\Lm_0}_\Lm [J]$, is related to $S_{\mathrm{int}}[\phi,\Lm]$
through \beq \label{stow} W^{\Lm_0}_\Lm [J]
=S_{\mathrm{int}}[P^{\Lm_0}_\Lm J,\Lm] -{1\over2}(J,P^{\Lm_0}_\Lm J). 
\eeq It satisfies the flow equation \beq \label{wflow} {\p
W^{\Lm_0}_\Lm\over\p\Lm} =-{1\over2}\int_q{\p R^{\Lm_0}_\Lm
(q^2)\over\p\Lm} \biggl\{{\delta^2W^{\Lm_0}_\Lm\over\delta J_q\,\delta
J_{-q}} -{\delta W^{\Lm_0}_\Lm\over\delta J_q} {\delta
W^{\Lm_0}_\Lm\over\delta J_{-q}}\biggr\}.  \eeq 

The cutoff effective action, $\Gamma^{\Lm_0}_\Lm$, is given by the
Legendre transform of $W^{\Lm_0}_\Lm [J]$, \beq
\Gamma^{\Lm_0}_\Lm[\varphi]= W^{\Lm_0}_\Lm[J]+(J,\varphi), \eeq where
$\varphi_q\equiv\delta W^{\Lm_0}_\Lm[J]/\delta J_{-q}$.  If one
further splits off a bare kinetic part of
$\Gamma^{\Lm_0}_\Lm[\varphi]$, \beq
\Gamma^{\Lm_0}_\Lm[\varphi]={1\over2} (\varphi,R^{\Lm_0}_\Lm\varphi)
+{\widetilde\Gamma}^{\Lm_0}_\Lm[\varphi], \eeq one obtains a flow
equation for ${\widetilde\Gamma}^{\Lm_0}_\Lm[\varphi]$, the generator
of 1PI vertex functions, of the form \beq \label{Legendre}
{\p{\widetilde\Gamma}^{\Lm_0}_\Lm\over\p\Lm} ={1\over2}\int_q{\p
R^{\Lm_0}_\Lm (q^2)\over\p\Lm} \biggl\{R^{\Lm_0}_\Lm (q^2)
+{\delta^2{\widetilde\Gamma}^{\Lm_0}_\Lm\over\delta\varphi_q\,
\delta\varphi_{-q}}\biggr\}^{-1}, \eeq with initial value \beq
\widetilde\Gamma^{\Lm_0}_{\Lm_0}[\varphi]=
S^{\Lm_0}_{\mathrm{int}}[\phi] \bigg|_{\phi=\varphi}.  \eeq

It is interesting to note that unlike the flow equations for
$S_{\mathrm{int}}$ and $\widetilde\Gamma^{\Lm_0}_\Lm$, which admit
well posed initial value problems in terms of the bare action
$S^{\Lm_0}_{\mathrm{int}}[\phi]$, the flow equation for
$W^{\Lm_0}_\Lm$ does not.  This is because the substitution
$\phi=P^{\Lm_0}_\Lm J$ used in eqs.~\re{rep2} and \re{stow} gives
$W^{\Lm_0}_{\Lm_0} [J]=0$ at $P^{\Lm_0}_{\Lm_0}=0$.  For this reason
we use eqs.~\re{polch} and \re{stow} rather than eq.~\re{wflow} for
the calculation of $W^{\Lm_0}_\Lm [J]$.  

\section{Rescaling} 

We recall
that Wilson's original renormalization group program consists of two
steps.  The first step, performed by the derivation given above,
integrates out the high momentum components of the field, generating
an equivalent effective action with reduced cutoff.  The second step
rescales the momenta back to the original cutoff scale and rescales
the fields so that the resulting equations are cutoff independent and
have a fixed point solution corresponding to the scale-invariant,
massless renormalized theory.  Bell and Wilson have shown
\cite{bellwilson} that for ``linear" renormalization group equations,
such as Wilson/Polchinski above, the field rescaling must be
appropriately carried out in terms of the scaling dimension of the
field or the resulting equations will not flow to a fixed point.  The
scale dependence of the cutoff function $P^{\Lm_0}_\Lm(q^2)$ must also
be chosen correctly.  Finally, we also transform to dimensionless
variables, which, while not really essential, is customary at this
stage.  

We parametrize our effective momentum scale by \beq t = \rm
{log} \: \bigg|{\Lm_0 \over \Lm }\bigg| , \eeq and our dimensionless
rescaled variables are \beq q' = {q \over \Lm} \eeq and \beq
\phi'_{q'} = \biggl({\Lm \over \Lm_0} \biggr)^{D-d_J}\phi_q \:
\Lm_0^{D-d_\phi}, \eeq where $d_J$ and $d_\phi$ are the scaling
dimensions of $J(x)$ and $\phi (x)$ respectively, \beq d_J
={1\over2}(D+2-\eta), \eeq and \beq d_\phi ={1\over2}(D-2+\eta), \eeq
where $\eta$ is the anomalous dimension.  A good choice for
$P^{\Lm_0}_\Lm(q^2)$ is \beq \label{cutoff}
P^{\Lm_0}_\Lm(q^2)=\Lm_0^{\eta -2}\biggl[\widetilde{P}(q^2/ \Lm_0^2)
-\biggl({\Lm \over \Lm_0} \biggr)^{2-\eta }\widetilde{P}(q^2/\Lm^2)
\biggr], \eeq where \[\widetilde{P}(q^2/\Lm^2)\to 1\hspace {0.3in}\rm
{for}\hspace {0.3in}q^2\ll \Lm^2\] \beq \widetilde{P}(q^2/\Lm^2)\to
0\hspace {0.3in}\rm {for}\hspace {0.3in}q^2\gg \Lm^2.  \eeq Finally,
by defining \beq S'_{\mathrm{int}}[\phi ',t]\equiv
S_{\mathrm{int}}[\phi (\phi '),\Lm (t)], \eeq we get \begin{eqnarray}
\label{rge} {\p S'_{\mathrm{int}}\over\p t} & = &D\,S'_{\mathrm{int}}-
\int_{q'} \phi'_{q'}\biggl[{1\over2}(D+2-\eta )+q'\cdot
\nabla'_{q'}\biggr]{\delta S'_{\mathrm{int}}\over\delta
\phi'_{q'}}\nonumber \\&& -\int_{q'}A(q') \biggl\{{\delta
S'_{\mathrm{int}}\over\delta \phi'_{q'}}{\delta S'_{\mathrm{int}}\over
\delta \phi'_{-q'}} -{\delta^2S'_{\mathrm{int}}\over\delta
\phi'_{q'}\,\delta \phi'_{-q'}}\biggr\}, \end{eqnarray} where
\begin{equation} A(q)=\biggl(1-{\eta \over 2}\biggr)\widetilde
P(q^2)-q^2{{\mathrm{d}}\widetilde P(q^2)\over {\mathrm{d}}(q^2)},
\end{equation} and the prime on $\nabla'_{q'}$ means that it ignores
the momentum conservation delta functions in ${\delta
S'_{\mathrm{int}}/\delta \phi'_{q'}}$.  

In the next section we will
show that our choice of rescaling and cutoff functions gives the
correct scaling laws at the critical point.  Here we note that if we
choose $\widetilde{P}(q^2)={\mathrm{e}}^{-2q^2},$ change variables to
$\sigma'_q={\mathrm{e}}^{q^2}\phi'_q$, and define \beq
{\cal{H}}[\sigma' ,t] \equiv -S'_{\mathrm{int}}[\phi'(\sigma'),t]-
{1\over 2}\int_q\sigma'_q\sigma'_{-q}, \eeq we get, after an
integration by parts and neglecting all primes, \begin{eqnarray} {\p
{\cal{H}}\over\p t} & = &\int_{q} \biggl({D\over2}\,\sigma_q+q\cdot \! 
\nabla_{q}\sigma_q\biggr){\delta {\cal{H}}\over\delta
\sigma_q}\nonumber \\&& +\int_{q}(1-{\eta \over2}+2q^2)
\biggl\{\sigma_q{\delta {\cal{H}}\over\delta \sigma_q}+{\delta
{\cal{H}} \over\delta \sigma_q}{\delta {\cal{H}}\over \delta
\sigma_{-q}} +{\delta^2{\cal{H}}\over\delta \sigma_q\,\delta
\sigma_{-q}}\biggr\}, \end{eqnarray} which is {\it exactly} Wilson's
equation \cite{wilsonkogut}.  Eq.~\re{rge} is more useful for our
purposes, however, and we will continue to work with it in what
follows.  

\section{Scaling Laws and Fluctuation Relations \\ for $N$-point Functions} 

In this section we derive a number of relations
for connected Green's functions based on their representation in terms
of the $t \to \infty \,\, (\Lm \to 0)$ solution to the ERG
eq.~\re{rge}.  The $UV$-regularized connected Green's functions are
determined from the generating functional \beq \label{wj} W^{\Lm_0}_0
[J] =\lim_{\Lm\to0}\biggl\{S_{\mathrm{int}}[P^{\Lm_0}_\Lm J,\Lm]
-{1\over2}(J,P^{\Lm_0}_\Lm J)\biggr\}, \eeq following eq.~\re{stow}. 
Let \begin{equation}
S'_{\mathrm{int}}[\phi',t]=\sum_{n=0}^{\infty}{1\over
n!}\int_{q'_1}\cdots \int_{q'_n}S'_n(q'_1,\ldots ,q'_n,t)
\phi'_{q'_1}\cdots \phi'_{q'_n}\delta^D (q'_1+\cdots +q'_n),
\end{equation} where the above is written in terms of our
dimensionless (primed) variables.  Rewriting this in terms of our
original (unprimed) variables, we get \begin{eqnarray}
S_{\mathrm{int}}[\phi,\Lm]&=&\Lm^D\sum_{n=0}^{\infty}{1\over
n!}\int_{q_1} \cdots \int_{q_n}S'_n(q{_1}\Lm^{-1},\ldots
,q{_n}\Lm^{-1},t)\nonumber \\&&\times
\Lm^{-nd_J}\Lm_0^{-n(d_J-d_\phi)} \phi_{q_1}\cdots \phi_{q_n}\delta^D
(q_1+\cdots +q_n).  \end{eqnarray} If we now substitute
$\phi_q=P_{\Lm}^{\Lm_0}(q^2)J_q$, we will then be able to study the
limit $\Lm \to 0$ of eq.~\re{wj}.  However, because
$P_0^{\Lm_0}(q^2)J_q=\Lm_0^{\eta -2}\widetilde{P}(q^2/\Lm_0^2)J_q$ is
finite we can substitute it directly in \re{wj} to get
\begin{eqnarray}
W^{\Lm_0}_0[J]&=&\lim_{\Lm\to0}\Lm^D\sum_{n=0}^{\infty} {1\over
n!}\int_{q_1}\cdots \int_{q_n}S'_n(q{_1}\Lm^{-1},\ldots ,q{_n}
\Lm^{-1},t)\Lm^{-nd_J}\nonumber \\&&\mbox{} \times
\widetilde{P}(q_1^2/\Lm_0^2)\cdots \widetilde{P}(q_n^2/\Lm_0^2)
J_{q_1}\cdots J_{q_n}\delta^D (q_1+\cdots +q_n)\nonumber \\
&&\mbox{}-{1\over2}\int_q \Lm_0^{\eta-2}\widetilde{P}(q^2/
\Lm_0^2)J_qJ_{-q}.  \end{eqnarray} In the same spirit we can now take
the limit $\Lm_0 \to \infty$, where $\Lm_0$ appears explicitly, to get
the infinite cutoff generating functional \beeq
W^{\infty}_0[J]&=&\lim_{\Lm_0\to\infty}\lim_{\Lm\to0}
\Lm^D\sum_{n=0}^{\infty}{1\over n!}\int_{q_1} \cdots
\int_{q_n}S'_n(q{_1}\Lm^{-1},\ldots ,q{_n}\Lm^{-1},t) \nonumber
\\&&\mbox{} \times \Lm^{-nd_J}J_{q_1}\cdots J_{q_n}\delta^D
(q_1+\cdots +q_n), \eeeq where the remaining limit $\Lm_0 \to \infty$
must be accompanied by appropriate fine tuning of the relevant bare
couplings to produce an acceptable renormalized theory
\cite{wilsonkogut,morsol,bagnuls}.  

The infinite cutoff connected Green's functions in a uniform external
field $j$ are 
\beq 
G_k(q_1,\ldots ,q_k,j)\delta^D (q_1+\cdots +q_k) 
=\mbox{}-{\delta^k\over \delta
J_{q_1}\cdots \delta J_{q_k}} W^{\infty}_0[J]\Bigm|_{J_{q_i}=(2\pi
)^Dj\delta^D (q_i)}, 
\eeq 
where $F_G(j)\equiv -G_0(j)$ is the Gibbs
free energy, the Legendre transform of the effective potential.  Thus,
\beeq 
\lefteqn{G_k(q_1,\ldots ,q_k,j)} \nonumber \\ & &
=\mbox{}-\lim_{\Lm_0\to\infty}\lim_{\Lm\to0}
\Lm^D\sum_{n=k}^{\infty}{1\over (n-k)!}S'_n(q{_1}\Lm^{-1},\ldots
,q{_k}\Lm^{-1},0,\ldots ,0,t)\nonumber \\&&\mbox{}\times
\Lm^{-kd_J}[j\Lm^{-d_J}]^{n-k}, 
\end{eqnarray} 
where the $G_k$ and
$S'_n$ are symmetric functions of their momentum arguments.  If we
define 
\beq \label{gktdef} 
\wG_k(q_1,\ldots ,q_k,x,t)\delta^D (q_1+\cdots +q_k)
\equiv \mbox{}-{\delta^k\over \delta \phi'_{q_1}\cdots \delta
\phi'_{q_k}}S'_{\mathrm{int}}[\phi',t]\,\Bigg|_{\phi'_{q_i}=(2\pi
)^Dx\delta^D (q_i)}, 
\eeq 
giving 
\beq \label{gktser} 
\wG_k(q_1,\ldots
,q_k,x,t)=\mbox{}-\sum_{n=k}^{\infty}{1\over (n-k)!}S'_n(q{_1}, \ldots
,q{_k},0,\ldots ,0,t)x^{n-k}, 
\eeq 
then 
\beq \label{gscale}
G_k(q_1,\ldots ,q_k,j)=\lim_{\Lm_0\to\infty}\lim_{\Lm\to0}
\Lm^{D-kd_J} \wG_k(q{_1}\Lm^{-1},\ldots ,q{_k}\Lm^{-1},j\Lm^{-d_J},t).
\end{equation} 
This equation allows us to derive fixed-point scaling laws and asymptotic 
behaviors as follows.

Let $S'^\ast_{\mathrm{int}}$ be a fixed-point solution of
eq.~\re{rge}: $\p S'^\ast_{\mathrm{int}}/\p t=0$.  The corresponding
$\wG^*_k$ are 
\beq 
\wG^\ast_k(q_1,\ldots,q_k,x)\delta^D (q_1+\cdots +q_k) 
= \mbox{} -{\delta^k\over \delta
\phi'_{q_1}\cdots \delta \phi'_{q_k}}S'^\ast_{\mathrm{int}}[\phi']\,
\Bigg|_{\phi'_{q_i}=(2\pi)^Dx\delta^D (q_i)}.  
\eeq 
In evaluating
eq.~\re{gscale} for such fixed-point solutions, the absence of
$t$-depen\-dence allows the limit ${\Lm_0\to\infty}$ to be trivially
taken, yielding 
\beq \label{gfixed} 
G^\ast_k(q_1,\ldots
,q_k,j)=\lim_{\Lm\to0} \Lm^{D-kd_J} \wG^\ast_k(q{_1}\Lm^{-1},\ldots
,q{_k}\Lm^{-1},j\Lm^{-d_J}).
\end{equation}
For this equation to have a nontrivial limit we must have $\wG^*_k\propto
\Lm^{-D+kd_J}$ as $\Lm\to0$.  We will consider this limit for two
cases.  First, for the case where $q_{i=1,\ldots,k}\neq 0$, we let
$x_i=q_i\Lm^{-1}$ and $y=j\Lm^{-d_J}$.  Then the $z_i\equiv
x_iy^{-{1/d_J}}=q_ij^{-{1/d_J}}$ are independent of $\Lm$, and
\begin{equation} x_1^{-d_J+{D\over k}}\cdots x_k^{-d_J+{D\over
k}}=\Lm^{-D+kd_J} q_1^{-d_J+{D\over k}}\cdots q_k^{-d_J+{D\over k}}. 
\end{equation} 
So for the limit $\Lm\to0$ to exist we must have
\begin{equation} \label{as1}
\wG^\ast_k(x_1,\ldots,x_k,y)
\mathop{\strut\longrightarrow}_{{x\to\infty
\atop y\to\infty} \atop z \,{\rm fixed}}
 {\cal{G}}_k(z_1,\ldots,z_k)x_1^{-d_J+{D\over k}}\cdots
x_k^{-d_J+{D\over k}},
\end{equation} 
where ${\cal{G}}_k$ is determined by solving the ERG 
equations.  Then
\beq \label{sc1} G^\ast_k(q_1,\ldots,q_k,j)=
 {\cal{G}}_k(z_1,\ldots,z_k)q_1^{-d_J+{D\over k}}\cdots
q_k^{-d_J+{D\over k}}.
\eeq 
This is the general (hyper)scaling law for $G_k$ on the critical
isotherm when $q_{i=1,\ldots,k}\neq 0$ \cite{fisher}.
For $k=2$, writing $G_2(q,-q,j)$ as
$G_2(q,j)$, we have
$G^\ast_2(q,j)={\cal{G}}_2(q/j^{2\over{D+2-\eta }})q^{\eta -2}$.   As this
expression generally is used to define the anomalous dimension
$\eta$, its derivation here validates our choice of rescaling
operations and cutoff functions. For the second case, where $j$ is
finite and $q_{i=1,\ldots,k}=0$, we must have
\beq \label{as2}
\wG^\ast_k(0,\ldots,0,y)\mathop{\strut\longrightarrow}_{y\to\infty}
 {\rm constant}\cdot y^{-k+{D\over {d_J}}}.
\eeq Then
\beq \label{sc2} G^\ast_k(0,\ldots,0,j)=
 {\rm constant}\cdot j^{-k+{D\over {d_J}}}.
\eeq

From the definition of $\wG_k$ we also have
\beq \label{gfr1}
\wG_k(q_1,\ldots ,q_{k-n},\overbrace{0,\ldots ,0}^{n\: \rm
zeroes},x,t)= {\p ^n \over \p x^n}\: \wG_{k-n}(q_1,\ldots
,q_{k-n},x,t),
\eeq which immediately gives
\beq \label{gfr2} G_k(q_1,\ldots ,q_{k-n},0,\ldots ,0,j)= {\p ^n
\over \p j^n}\: G_{k-n}(q_1,\ldots ,q_{k-n},j).
\eeq These relations will be important in what follows.  They
are a generalization of the so-called ``fluctuation-dissipation
theorem" or ``linear response theorem" of statistical mechanics, which
relates fluctuations in a thermodynamic average to a corresponding
susceptibility (linear response)
\cite{fisher,binney}.  The case most often encountered in statistical
mechanics is for
$k=n=2$, which gives the well-known relation between the 2-point
cumulant and order parameter susceptibility,
\beq G_2(0,j)=\p \varphi _0/\p j,
\eeq where $\varphi _0=\p G_0/\p j$.  We will refer to
relations~\re{gfr1} and \re{gfr2} as {\it fluctuation relations} in
what follows.

\section{Flow Equations for $N$-point Functions}

The flow equations for the $\wG_k$ can now be derived by applying the
definition~\re{gktdef} to the flow equation~\re{rge}.
Using~\re{gktdef} we can write
\beq
\delta^D (q_1+\cdots +q_k){\p \wG_k\over \p t}=-\biggl({\delta^k\over
\delta
\phi'_{q_1}\cdots \delta
\phi'_{q_k}}\;{\p S'_{\mathrm{int}}\over \p
t}\biggr)_{\phi'_{q_i}=(2\pi)^Dx\delta^D (q_i)}.
\eeq Then, by substituting the r.h.s. of eq.~\re{rge} for $\p
S'_{\mathrm{int}}/\p t$ and carrying out the indicated operations, we
get (suppressing
$q$, $x$, and $t$ arguments when not necessary for clarity)
\beeq \label{gflow1} {\p \wG_0\over \p t}&=&D \wG_0-d_J\, x{\p
\wG_0\over
\p x}+A(0)\biggl({\p \wG_0\over \p x}\biggr)^2+\int_q A(q)\wG_2(q)
\nonumber \\ {\p \wG_{k>0}\over \p t}&=&(D-kd_J)\wG_k-d_J\, x{\p
\wG_k\over
\p x}-\sum_{i=1}^kq_i\cdot \nabla_{q_i}\wG_k \nonumber \\
&&+\sum_{j=0}^k\sum_{I_j\subset Z_k} \biggl\{A(q_{i_1}+\cdots
+q_{i_j})\widetilde G_{j+1}(-q_{i_1}-\cdots -q_{i_j},q_{i_1},\ldots
,q_{i_j}) \nonumber \\ &&\times \wG_{k-j+1}(-q_{i_{j+1}}-\cdots
-q_{i_k},q_{i_{j+1}},\ldots,q_{i_k})\biggr\} \nonumber \\
&&+\int_{q'}A(q')\wG_{k+2}(q',-q',q_1,\ldots ,q_k),
\eeeq where $I_j$ stands for the set of indices $\{i_1,\ldots
,i_j\}$, $Z_k$ stands for the set of consecutive integers $\{1,\ldots
,k\}$, $\sum_{I_j\subset Z_k}$ stands for the sum over the
$\bigl({k\atop j}\bigr)$ different sets, $I_j$, of indices chosen
from the set $Z_k$, and the set $\{i_{j+1},\ldots ,i_k\}$ is the
complementary set
$Z_k-I_j$.   We note that in the equation for
$\p
\wG_k/\p t$ we have
$q_1+\cdots +q_k=0$ from the $\delta $-function of~\re{gktdef}.
Using this together with eq.~\re{gfr1} we can rewrite
eqs.~\re{gflow1} as
\beeq \label{gflow2} {\p \wG_0\over \p t}&=&D \wG_0-d_J\, x{\p
\wG_0\over
\p x}+A(0)\biggl({\p \wG_0\over \p x}\biggr)^2+\int_q A(q)\wG_2(q)
\nonumber \\ {\p \wG_{k>0}\over \p t}&=&(D-kd_J)\widetilde
G_k+\biggl(2A(0){\p \wG_0\over \p x}-d_J\, x\biggr){\p
\wG_k\over
\p x}-\sum_{i=1}^kq_i\cdot \nabla_{q_i}\wG_k \nonumber \\
&&+\sum_{j=1}^{k-1}\sum_{I_j\subset Z_k} \biggl\{A(q_{i_1}+\cdots
+q_{i_j})\wG_{j+1}(-q_{i_1}-\cdots -q_{i_j},q_{i_1},\ldots ,q_{i_j})
\nonumber \\ &&\times \wG_{k-j+1}(-q_{i_{j+1}}-\cdots
-q_{i_k},q_{i_{j+1}},\ldots,q_{i_k})\biggr\} \nonumber \\
&&+\int_{q'}A(q')\wG_{k+2}(q',-q',q_1,\ldots ,q_k),
\eeeq where we have explicitly pulled out the first and last terms
from the double sum.

The first four equations are
\beeq \label{gflowk} {\p \wG_0\over \p t}&=&D \wG_0-d_J\, x{\p
\wG_0\over
\p x}+A(0)\biggl({\p \wG_0\over \p x}\biggr)^2+\int_q A(q)\wG_2(q)
\nonumber \\ {\p \wG_2\over \p t}&=&(D-2d_J)\wG_2+\biggl(2A(0){\p
\wG_0\over \p x}-d_J\, x\biggr){\p\wG_2\over
\p x}-q\cdot \nabla_{q}\wG_2 \nonumber \\
&&\mbox{}+2A(q)\wG_2^2+\int_{q'}A(q')\wG_4(q',-q',q,-q)
\nonumber
\\  {\p \wG_3\over \p t}&=&(D-3d_J)\wG_3+\biggl(2A(0){\p
\wG_0\over \p x}-d_J\, x\biggr){\p\wG_3\over
\p x}-\sum_{i=1}^3q_i\cdot \nabla_{q_i}\wG_3 \nonumber \\
&&\mbox{}+2\biggl[\sum_{i=1}^3A(q_i)\wG_2(q_i)
\biggr]\wG_3(q_1,q_2,q_3)
+\int_{q'}A(q')\wG_5(q',-q',q_1,q_2,q_3)\nonumber \\  {\p \wG_4\over
\p t}&=&(D-4d_J)\wG_4+\biggl(2A(0){\p \wG_0\over
\p x}-d_J\, x\biggr){\p \wG_4\over
\p x}-\sum_{i=1}^4q_i\cdot \nabla_{q_i}\wG_4 \nonumber \\
&&\mbox{}+2\biggl[\sum_{i=1}^4A(q_i)\wG_2(q_i)
\biggr]\wG_4(q_1,q_2,q_3,q_4)
\nonumber
\\ &&+\sum_{I_2\subset
Z_4}A(q_{i_1}+q_{i_2})\wG_3(-q_{i_1}-q_{i_2},q_{i_1},q_{i_2})
\wG_3(-q_{i_3}-q_{i_4},q_{i_3},q_{i_4}) \nonumber \\
&&+\int_{q'}A(q')\wG_6(q',-q',q_1,\ldots ,q_4).
\eeeq We note that $\wG_1(q,x,t)=\wG_1(0,x,t)=\p \wG_0(x,t)/\p x$ due
to the $\delta^D (q)$ in the definition of $\wG_1(q,x,t)$, so we
don't need to write a separate equation for it.  Also, in response
to the $\delta $-function with
$\wG_2$, we write
$\wG_2(q,-q,x,t)$ as $\wG_2(q,x,t)$.

A natural procedure for solving the above hierarchy of equations is
to truncate at a certain level by setting all higher order connected
Green's functions equal to zero.  To lowest order we get an equation
for $\wG_0$ that has been studied long ago by Green, Gunton, and
coworkers \cite{green,gunton,dee}.  Green
\cite{green} derived this equation via a steepest descent
approximation to Wilson's functional integral form of his ERG
equation \cite{wilsonkogut}.  He showed that, with appropriate
rescaling to ensure analyticity, mean field exponents are obtained
\cite{green}.  It was later shown that the equations reproduce more
of the mean field phenomenology, including metastable and unstable
branches in the free energy \cite{dee} and a spinodal fixed point
\cite{gunton}.  The higher order equations thus reflect systematic
corrections to mean field behavior when successively higher order
correlations are taken into account.

A perspective on the higher order equations may be gained by
realizing that they have already been studied for the zero-field case
($x=0$) in $D=4-\epsilon $ dimensions by Green and Shukla
\cite{shuklagreen}.  Because the equations are first-order
quasi-linear they can be solved exactly in $\epsilon$-expansion,
reproducing the standard results for the critical exponents.  The
$\epsilon$-expansion provides a natural truncation scheme for the
hierarchy of equations, as $\wG_{2n}$ is of order $\epsilon^{n-1}$
for $n\geq 2$.  As the $\epsilon$-expansion is known to be
asymptotic, we do not expect similar truncation schemes without the
benefit of $\epsilon$ as a small parameter to yield any improvement
in convergence.  Nor, due to continuity of the equations in
$x$, do we expect different convergence properties for finite values
of $x$.  For this reason we do not expect the obvious truncation
scheme to yield a convergent method of successive approximation.
This is not to say, however, that low orders of approximation could
not be useful.

\section{Momentum Cluster Flow Equations}

It is possible to improve upon the above truncation scheme by noting
that, according to the fluctuation relations~\re{gfr1}, each $\wG_k$
contains derivatives of $\wG_{k'<k}$ within it.  Thus the truncation
scheme discussed above needlessly throws away vital information.  A
truncation scheme that retains this information from higher order
correlations can be constructed as follows.

Due to the momentum conserving $\delta $-function in the
definition of
$\wG_k$, $\wG_k$ is a function of only $(k-1)$ independent momenta,
and we will write it henceforth as $\wG_k(q_1,\ldots
,q_{k-1},x,t)$, with the understanding that the omitted
$q_k$ takes the value $-q_1-\cdots -q_{k-1}$.  Defining the projection
operator
${\cal P}_{q_i}\wG_k(q_1,\ldots ,q_{k-1},x,t)\equiv
\wG_k(q_1,\ldots ,q_i\! =\! 0,\ldots ,q_{k-1},x,t)$,
we define, for
$k\geq 2$,
\beq
\hG_{k\geq 2}(q_1,\ldots ,q_{k-1},x,t)\equiv
\biggl[\,\prod_{i=1}^{k-1}(1-{\cal P}_{q_i})\biggr]\wG_k(q_1,\ldots
,q_{k-1},x,t).
\eeq   We also define $\hG_0(x,t)\equiv \wG_0(x,t)$ and
$\hG_1(x,t)\equiv 0$.  Clearly,
$\hG_{k\geq 2}=0$ if any of its
$(k-1)$ momenta are zero.  We call
$\hG_k$ the {\it k-point momentum cluster} because {\it all} of its
momentum variables are present in each term of its
power series expansion.  By expanding the product and making use of
the fluctuation relations~\re{gfr1} we obtain
\beeq \label{mclinv}
\lefteqn{\hG_{k\geq 2}(q_1,\ldots ,q_{k-1},x,t)}
\nonumber \\ & & = \sum_{n=0}^{k-2}\:(-1)^n\sum_{I_{k-1-n}\atop
\subset
\:Z_{k-1}}{\p^n\over \p x^n}\wG_{k-n}(q_{i_1},\ldots
,q_{i_{k-1-n}},x,t) \mbox{}+(-1)^{k-1}{\p^k\over \p x^k}\wG_0.
\eeeq
In a similar manner, using the identity
 $1={\cal P}_{q_i}+(1-{\cal P}_{q_i})$, we find the inverse relation
\beeq \label{mcl}
\lefteqn{\wG_{k\geq 2}(q_1,\ldots ,q_{k-1},x,t)}
\nonumber \\ & & = \sum_{n=0}^{k-2}\:\sum_{I_{k-1-n}\atop \subset
\:Z_{k-1}}{\p^n\over \p x^n}\hG_{k-n}(q_{i_1},\ldots
,q_{i_{k-1-n}},x,t)+{\p^k\over \p x^k}\hG_0.
\eeeq We call eq.~\re{mcl} the {\it momentum cluster decomposition}
of $\wG_k$.

Using these relations together with the flow equations for the
$\wG_k$~\re{gflow2} we can write flow equations for the $\hG_k$ as
follows.  Again we suppress unnecessary arguments for the sake of
clarity.  Define
$F_k[\wG_0,\ldots ,\wG_k]$ by writing the flow equations~\re{gflow2}
as
\beq \label{gflowc} {\p \wG_k\over \p t}=F_k[\wG_0,\ldots
,\wG_k]+\int_{q'}A(q')\wG_{k+2}(q',-q',q_1,\ldots
,q_{k-1}).
\eeq Then
\beeq \label{mclflow1} {\p \hG_0\over \p
t}&=&F_0[\wG_0]+I_A\:{\p^2\hG_0\over \p x^2} +\int_{q}A(q)\hG_2(q)
\nonumber \\ {\p \hG_{k\geq 2}\over \p
t}&=&\sum_{n=0}^{k-1}\:(-1)^n\sum_{I_{k-1-n}\atop \subset
\:Z_{k-1}}{\p^n\over \p x^n}\,F_{k-n}[\wG_0,\ldots
,\wG_{k-n}]+I_A\:{\p^2\hG_k\over \p x^2}
\nonumber \\ & &+\int_{q'}A(q')\hG_{k+2}(q',-q',q_1,\ldots ,q_{k-1}),
\eeeq where $I_A\equiv \int_{q}A(q)$ and $F_1[\wG_0,\wG_1]=\p
F_0[\wG_0]/\p x$.  To complete the derivation of the flow equations
for $\hG_k$ we must reexpress the arguments of
$F_{k-n}$ in terms of
$\hG_{k'\leq k-n}$ via eq.~\re{mcl}, but already in the above form we
can see our most significant results, which we will describe shortly.

The first four {\it momentum cluster flow equations} are
\beeq \label{mcflowk} {\p \hG_0\over \p t}&=&D \hG_0-d_J\,x\pdof
{\hG_0}+A(0)\biggl(\pdof {\hG_0}\biggr)^2+I_A\:\pd 2{\hG_0} +\int_q
A(q)\hG_2(q)
\nonumber \\  {\p \hG_2\over \p t}&=&(D-2d_J)\hG_2+\biggl(2A(0)\pdof
{\hG_0}-d_J\, x\biggr)\pdof {\hG_2}-q\cdot \nabla_{q}\hG_2 
-2A(0)\biggl(\pd 2{\hG_0}\biggr)^2\nonumber
\\ &&\mbox{}+2A(q)\biggl(\hG_2+\pd 2{\hG_0}\biggr)^2 
+I_A\:\pd 2{\hG_2} +\int_{q'}A(q')\hG_4(q',-q',q)
\nonumber \\   {\p \hG_3\over \p t}&=&(D-3d_J)\hG_3+\biggl(2A(0)\pdof
{\hG_0}-d_J\, x\biggr)\pdof {\hG_3}-\sum_{i=1}^2q_i\cdot
\nabla_{q_i}\hG_3 \nonumber \\ & &\mbox{}+2A(0)\pd 2{\hG_0}\biggl[\pd
3{\hG_0} -\sum_{i=1}^2\pdof {\hG_2(q_i)}\biggr] 
\nonumber \\  & &\mbox{}+2\sum_{\{i,j\}\atop
\epsilon \,C_2}A(q_i)\biggl[\hG_2(q_i)
+\pd 2{\hG_0}\biggr]\biggl[\hG_3(q_1,q_2)+\pdof {\hG_2(q_j)} -\pdof
{\hG_2(q_i)}-\pd 3{\hG_0}\biggr]
\nonumber \\ & &\mbox{}+2A(q_1+q_2)\biggl[\hG_2(q_1+q_2)+\pd
2{\hG_0}\biggr]
\biggl[\hG_3(q_1,q_2)+\sum_{i=1}^2\pdof {\hG_2(q_i)}
+\pd 3{\hG_0}\biggr] \nonumber \\ & &\mbox{} +I_A\pd
2{\hG_3}+\int_{q'}A(q')\hG_5(q',-q',q_1,q_2)
\nonumber \\  {\p \hG_4\over \p t}&=&(D-4d_J)\hG_4+\biggl(2A(0)\pdof
{\hG_0}-d_J\, x\biggr)\pdof {\hG_4}-\sum_{i=1}^3q_i\cdot
\nabla_{q_i}\hG_4 \nonumber \\ & &\mbox{}-2A(0)\biggl[\pd
2{\hG_0}\sijk \pdx \hG_3(q_j,q_k)-\pd 3{\hG_0}\sum_{i=1}^3\pdof
{\hG_2(q_i)} +\pd 4{\hG_0}\,\pd 2{\hG_0}\biggr] 
\nonumber \\  & &\mbox{}+2\sijk A(q_i)\biggl\{\biggl[\hG_2(q_i)+\pd
2{\hG_0}\biggr]\biggl[\hG_4(q_1,q_2,q_3)
\nonumber \\ &&\mbox{}+\pdx \hG_3(q_j,q_k)-\pdx \hG_3(q_i,q_j)-\pdx
\hG_3(q_i,q_k)+\pd 2{\hG_2(q_i)}
+\pd 4{\hG_0}\biggr]\nonumber \\   & &\mbox{}-\biggl[\pd
2{\hG_2(q_i)}+\pd 3{\hG_0}\biggr]\biggl[\hG_3(q_i,q_j)+\hG_3(q_i,q_k)
+\pdof {\hG_2(q_j)}+\pdof
{\hG_2(q_k)}\biggr]\biggr\}\nonumber \\  & &\mbox{}+2\sijk A(q_j+q_k)
\biggl\{\biggl[\hG_3(-q_j-q_k,q_i)+\pdof {\hG_2(q_j)}+\pdof
{\hG_2(q_k)}+\pd 3{\hG_0}\biggr] \nonumber \\  & &\mbox{}\times
\biggl[\hG_3(q_j+q_k,q_i)+\pdof
{\hG_2(q_i)}\biggr]
-\biggl[\hG_2(q_j+q_k)+\pd 2{\hG_0}\biggr]
\nonumber \\  & &\mbox{}\times \biggl[\pdx \hG_3(q_j,q_k)+\pd
2{\hG_2(q_j)}+\pd 2{\hG_2(q_k)}+\pd 4{\hG_0}\biggr]\biggr\}
\nonumber \\  & &\mbox{}+2A(q_1+q_2+q_3)\biggl[\hG_2(q_1+q_2+q_3)+\pd
2{\hG_0}\biggr]\biggl[\hG_4(q_1,q_2,q_3) \nonumber \\ &
&\mbox{}+\sijk \pdx \hG_3(q_j,q_k)+\sum_{i=1}^3\pd 2{\hG_2(q_i)}+\pd
4{\hG_0}\biggr]
\nonumber \\  & &\mbox{}+I_A\pd
2{\hG_4}+\int_{q'}A(q')\hG_6(q',-q',q_1,q_2,q_3),
\eeeq where $C_k$ stands for the cyclic permutations of $\{1,\ldots
,k\}$.

The above equations display many features which appear to give them
distinct advantages over the standard $N$-point flow
equations~\re{gflowk}.  By comparing eqs.~\re{mcflowk} with
eqs.~\re{gflowk} we see that the number of terms in the equation for
$\p \hG_k /\p t$ increases dramatically with $k$ relative to its
``unclustered" counterpart \re{gflowk}.  This is also directly
apparent in comparing eqs.~\re{mclflow1} and eqs.~\re{gflowc}.  The
new terms are those which are discarded in the standard truncation
approach to the flow equations \re{gflowc}, and their presence in
\re{mclflow1} is one measure of the improvement of the momentum
cluster approach.  Because more terms are ``rescued" as $k$ increases
this may offer hope for convergence of the approximation procedure.

A more direct measure of the amount of information retained by the
two sets of equations after truncation is obtained by comparing
their integral terms, which contain the contribution from higher
order functions which is lost when the equations are truncated.
Their relative importance can be estimated by noting that, according
to \re{as1}, the $\wG_k^*$ go to zero as an inverse power of the
$q_{i=1,\ldots ,k}$ as the $q_i$ go to infinity.  As they generally
are nonzero for
$q_i=0$, we can assume their weight to be largest in the region near
the origin and hence make a major contribution to the integral term
in eqs.~\re{gflowc}.  This is because the function $A(q)$ is
typically concentrated near the origin as well, generally chosen to
behave as ${\mathrm{e}}^{-aq^2}$ for large $q$.  In contrast, the
functions $\hG_k$ have had, by definition, their zero point values
subtracted away, leaving functions that are concentrated away from
the origin, tending to nonzero values at infinity.  These functions
will therefore make a much smaller contribution to the integrals of
eqs.~\re{mclflow1} and, hopefully, be much less missed upon
truncation.

Another significant difference between the two sets of equations is
that the flow equations for $\hG_k$ are parabolic in $x$ rather than
first order, due to the presence of the $I_A\p^2 \hG_k/\p x^2$ term
in the equation for $\p \hG_k /\p t$.   This causes a dramatic change
in their solution structure relative to the flow equations for the
$\wG_k$.  This is already apparent in that the new lowest order of
approximation is the LPA approximation.  This represents a
significant improvement over its mean field counterpart of
eqs.~\re{gflowk}, both in terms of a more accurate calculation of
critical exponents and description of universality classes and in
terms of (presumably) avoiding the thermodynamically unstable
branches of the mean field free energy.  If the higher order
equations of \re{mcflowk} represent as much of an improvement over
their eq.~\re{gflowk} counterparts as the LPA approximation is over
the mean field approximation, we can expect these equations to prove
very useful.

Another consequence of the parabolic nature of the flow
equations~\re{mcflowk} is that they cannot be solved directly in the
$x=0$ limit, as could eqs.~\re{gflowk}.  Thus, the full
$x$-dependence of the flow equations will play an essential role in
their approximation behavior.

Of course, while the above points suggest the possibility of much
improved approximations based on truncations of the momentum cluster
flow equations, the convergence of the truncation scheme remains an
open question.  Yet, a hopeful indication that convergence may not be
out of reach lies in the fact that the momentum cluster flow
equations are essentially an expansion of the ERG equation in terms
of the {\it number} of momenta which couple the Fourier transformed
field variables in the effective action.  In a lattice real-space
formulation this corresponds to an expansion of the effective lattice
action in terms of local interactions ordered according to the {\it
number} of interacting sites (of arbitrary separation).  Thus, we
begin to make contact, perhaps with improvements, with the apparently
convergent approximation scheme discussed in our Introduction.

\section{Concluding Discussion}

We close with some observations about future directions.  The
prospect for solving the lower order equations numerically looks
quite good, as algorithms and approaches for solving PDEs are readily
available \cite{numrec} and have already begun to be applied to the
solution of approximate RG equations
\cite{nvect,ellbound,wettsol}.  These approaches should put solution
of the $\hG_4$ equation within reach, at least in so far as it
contributes to the equation for
$\hG_2$, as for that purpose only $\hG_4(q_1,-q_1,q_2,x,t)$ is
required.  Thus, at this level of truncation, only four variables,
$q_1^2, q_2^2, (q_1+q_2)^2,$ and $x$, are needed to solve for the
fixed-point solution and critical exponents.  To determine the full
momentum dependence of $\hG_4$, or the higher order corrections
needed for the empirical study of convergence, will probably require
further approximation \cite{ellbound}.

Questions of convergence aside, the momentum cluster approach
provides a possible advantage over the usual formulation in that it
works directly with constant-field connected Green's functions rather
than the interactions contributing to the RG effective action.  Since
the field is constant, it has no momentum index, considerably
simplifying the description of momentum dependence when the approach
is extended to vector models and gauge theories.  This suggests that
equations analogous to the $\hG_2$ equation may be a useful starting
approximation for the investigation of the space of renormalizable
gauge theories, just as the LPA approximation was for scalar and
vector models \cite{nvecLPA,morsol}.

The momentum cluster approach can also be applied to the Legendre
transformed ERG equation~\re{Legendre} \cite{wett,morijmp} and
possibly to sharp-cutoff versions of the equation
\cite{morijmp,wegnerhoughton,momscale}, as well.  In this paper we
limited ourselves to the Wilson/Polchinski version because it had the
simplest form of nonlinear terms.  More generally, one might consider
whether the momentum cluster approach might have other applications,
as many areas of physics make use of hierarchies of equations that
begin with the mean field approximation and involve corrections from
higher order correlations.  If these equations can be represented in
a generating functional format, the same fluctuation relations and
momentum cluster decomposition might well apply, giving access to a
new hierarchy of equations that no longer begins at a mean field
level of approximation.

\section*{Acknowledgments}

I would like to thank Tim Morris for inspiration and encouragement,
as well as for many useful and stimulating discussions.  Thanks also
to Valery Kholodnyi for patient and timely mathematical instruction,
to David Golner and Sunil Rawal for comments on the manuscript, and 
to an anonymous referee for noting an error in Section 4 regarding the 
infinite cutoff limit.

\end{document}